\documentclass[showpacs,preprintnumbers,nofootinbib,amsmath,amssymb,preprint]{revtex4}
\usepackage[dvips]{color}
\usepackage{graphicx}

\newcommand{\eV}{{\text{eV}}}
\newcommand{\GeV}{{\text{GeV}}}
\newcommand{\eVGeV}{{\text{eV$\cdot$GeV}}}
\newcommand{\MNS}{{\text{MNS}}}
\newcommand{\BR}{{\text{BR}}}
\newcommand{\magic}{{\text{mgc}}}
\newcommand{\vm}{v_\Delta m_{H^{\pm\pm}}}
\newcommand{\meee}{{\mu \to \bar{e}ee}}
\newcommand{\meg}{{\mu \to e\gamma}}
\newcommand{\tmmm}{{\tau \to \bar{\mu}\mu\mu}}
\newcommand{\tmee}{{\tau \to \bar{\mu}ee}}

\begin{document}
\preprint{IC/2009/076}
\title{Constraints from muon $g-2$ and LFV processes
in the Higgs Triplet Model}
\author{Takeshi Fukuyama}
\email{fukuyama@se.ritsumei.ac.jp}
\affiliation{Department of Physics and R-GIRO,
Ritsumeikan University, Kusatsu, Shiga,
525-8577, Japan}
\author{Hiroaki Sugiyama}
\email{hiroaki@fc.ritsumei.ac.jp}
\affiliation{Department of Physics and R-GIRO,
Ritsumeikan University, Kusatsu, Shiga,
525-8577, Japan}
\author{Koji Tsumura}
\email{ktsumura@ictp.it}
\affiliation{The Abdus Salam ICTP of UNESCO and IAEA,
Strada Costiera 11, 34151 Trieste, Italy}
\begin{abstract}
 Constraints from the muon anomalous magnetic dipole moment
and lepton flavor violating processes are translated
into lower bounds on $\vm$ in the Higgs Triplet Model
by considering correlations through
the neutrino mass matrix.
 The discrepancy of the sign of the contribution 
to the muon anomalous magnetic dipole moment
between the measurement and the prediction in the model
is clarified.
 It is shown that
$\mu \to e\gamma$, $\tau$ decays (especially, $\tmee$),
and the muonium conversion can give a more stringent
bound on $\vm$ than the bound from $\meee$
which is expected naively to give
the most stringent one.
\end{abstract}
\maketitle

\section{Introduction}

 In the standard model of the particle physics (SM),
neutrinos are massless particles due to the absence of
right-handed neutrinos $\nu_R$.
 The simplest way to give masses to three neutrinos
is to add three $\nu_R$ similarly to other fermions,
which corresponds to six additional particles
(three $\nu_R$ and three $\overline{\nu_R}$)
to the SM\@.
 In the Higgs triplet model (HTM)~\cite{HTM,Schechter:1980gr}
which we deal with in this article,
a complex $SU(2)_L$ triplet scalar with the hypercharge $Y=2$
is introduced to the SM in order to have neutrino masses.
 This model can be regarded as one of the simplest extension of the SM
because the number of new particles is six in this model also.

 The triplet Higgs boson field with hypercharge $Y=2$ 
can be parameterized by
\begin{align}
\Delta
\equiv
 \begin{pmatrix}
  \Delta^+/\sqrt2                     & \ \Delta^{++}\\
  \frac{v_\Delta^{}}{\sqrt2}+\Delta^0 & \ -\Delta^+/\sqrt2
 \end{pmatrix},
\end{align}
where $v_\Delta^{}$ is the vacuum expectation value (VEV) of 
the triplet Higgs boson. 
 The constraint on the rho parameter,
$\rho_0 = 1.0004^{+0.0027}_{-0.0007}$ at $2\sigma$~CL
(page 137 of \cite{Amsler:2008zzb}),
gives an upper limit $v_\Delta/v \lesssim 0.01$
where $v = 246\,\GeV$ is the VEV of the doublet Higgs field,
which corresponds to $v_\Delta^{}\lesssim 3\,\GeV$. 
There is no stringent bound from quark sector on triplet Higgs bosons 
because they do not couple to quarks.
 The interaction of the Higgs triplet
with lepton doublets $L_\ell \equiv (\nu_{\ell L}, \ell_L)^T$
$(\ell = e,\mu,\tau)$ is given by
\begin{align}
{\mathcal L}_{\text{triplet-Yukawa}}
 &= h_{\ell\ell^\prime}\,
    \overline{L_\ell^c}\,i\tau_2\Delta L_{\ell^\prime} + \text{H.\ c.}
\end{align}
 The symmetric matrix
$h_{\ell\ell^\prime}$ is coupling strength,
$\tau_i(i=1$--$3)$ denote the Pauli matrices,
and $\overline{L_\ell^c}
\equiv \left( \nu_{\ell L}^TC,\, \ell_L^TC \right)$
with the charge conjugation operator $C$.

 The coupling $h_{\ell\ell^\prime}$ has a direct relation
to the neutrino mass matrix $m_{\ell\ell^\prime}$
in the flavor basis through $v_\Delta$ as
\begin{equation}
h_{\ell\ell^\prime}
=
 \frac{1}{\sqrt{2} v_\Delta}
 \left(
  U_\MNS^\ast\,
  \text{diag}(m_1, m_2 e^{-i\varphi_1}, m_3 e^{-i\varphi_2})\,
  U_\MNS^\dagger
 \right)_{\ell\ell^\prime}
\equiv
 \frac{1}{\sqrt{2} v_\Delta} m_{\ell\ell^\prime}.
\label{eq:hM}
\end{equation}
 The mass eigenvalues $m_i$ are taken to be real positive values.
 We define $\Delta m^2_{ij} \equiv m_i^2 - m_j^2$
and refer to the case of $\Delta m^2_{31} > 0$ ($\Delta m^2_{31} < 0$)
as the normal (inverted) mass ordering.
 Here neutrinos are required to be Majorana particles%
\footnote
{
 In general,
even if the Higgs triplet exists,
neutrinos can be Dirac particles
by adding $\nu_R$ also and requiring lepton number conservation
which results in $v_\Delta = 0$.
 In the HTM we use,
neutrino masses are assumed to be given solely by $v_\Delta$
and neutrinos are Majorana particles by definition.
},
and $\varphi_1$ and $\varphi_2$ are the Majorana phases%
~\cite{Schechter:1980gr,Bilenky:1980cx}
defined in an interval of $[0,2\pi)$.
 The Maki-Nakagawa-Sakata matrix~\cite{Maki:1962mu}
of the neutrino mixing%
\footnote
{
 We took the definition of the mixing
$\nu_\ell = \sum_i U_{\ell i} \nu_i$
according to page 517 of \cite{Amsler:2008zzb}
although another definition
$\nu_\ell = \sum_i U^\ast_{\ell i} \nu_i$
is used for example, in \cite{Akeroyd:2009nu}
and on page 163 of \cite{Amsler:2008zzb}.
 In latter definition,
we need to take complex conjugate
in the middle equation of (\ref{eq:hM}).
}
is parameterized as
\begin{equation}
U_\MNS
\equiv
\left(
\begin{array}{ccc}
 c_{12}c_{13}
  & s_{12}c_{13}
  & s_{13}e^{-i\delta} \\
 -s_{12}c_{23}-c_{12}s_{23}s_{13}e^{i\delta}
  & c_{12}c_{23}-s_{12}s_{23}s_{13}e^{i\delta}
  & s_{23}c_{13} \\
 s_{12}s_{23}-c_{12}c_{23}s_{13}e^{i\delta}
  & -c_{12}s_{23}-s_{12}c_{23}s_{13}e^{i\delta}
  & c_{23}c_{13}  
\end{array}
\right)
\,,
\end{equation} 
where $s_{ij}\equiv\sin\theta_{ij}$ and $c_{ij}\equiv \cos\theta_{ij}$,
and $\delta$ is the Dirac phase.
 The ranges are chosen as $0\leq\theta_{ij}\leq\pi/2$
and $0\leq\delta<2\pi$.
 According to current constraints from
neutrino oscillation experiments~\cite{solar,atm,acc,Apollonio:2002gd},
we use the following values in this article
\begin{eqnarray}
&&
 \Delta m^2_{21} = 7.6\times 10^{-5}\,\eV^2, \ \
 |\Delta m^2_{31}| = 2.4\times 10^{-3}\,\eV^2,
\label{param1}\\
&&
 \sin^2{2\theta_{12}} = 0.87, \ \
 \sin^2{2\theta_{23}} = 1,
\label{param2}\\
&&
 \sin^2{2\theta_{13}} < 0.14
\label{param3}.
\end{eqnarray}
 The absolute scale of the neutrino mass is
constrained by tritium beta decay measurements
as $m_\nu \leq 2.3\,\eV$~(95\%~CL)~\cite{Kraus:2004zw}
and by cosmological observations
as $\sum m_i < 0.61\,\eV$~(95\%~CL)
or $\sum m_i < 1.3\,\eV$~(WMAP only, 95\%~CL)~\cite{WMAP5}.

 The HTM has seven physical Higgs bosons
which are
two CP-even neutral bosons $h^0$ (lighter) and $H^0$ (heavier),
a CP-odd neutral one $A^0$,
a pair of singly charged bosons $H^\pm$,
and a pair of doubly charged bosons $H^{\pm\pm}$.
 These Higgs bosons contribute to
many lepton flavor violating (LFV) processes.
 Experimental searches for $\meee$ etc.\
put upper bounds on $|h_{\mu e}| |h_{ee}|/m_{H^{\pm\pm}}^2$ etc.\
(See e.g.\ \cite{Abada:2007ux}),
where $m_{H^{\pm\pm}}$ is the mass of $H^{\pm\pm}$.
 The couplings $h_{\ell\ell^\prime}$ are,
however, not free in the HTM
because they relate directly to
the neutrino mass matrix $m_{\ell\ell^\prime}$
as shown in (\ref{eq:hM}).
 Previous works for dependences of LFV processes
on the parameters in $m_{\ell\ell^\prime}$
can be found in \cite{Chun:2003ej,Kakizaki:2003jk,Akeroyd:2009nu}.
 In this article,
we consider in detail the correlation
of upper bounds on $|h_{ij}^\ast h_{kl}|/m_{H^{\pm\pm}}^2$
from new physics searches
and deal with them as lower bounds on $\vm$.

\section{Lower bound on $v_\Delta m_{H^{\pm\pm}}$}

\subsection{Constraint from the Muon Anomalous Magnetic Dipole Moment}

 Let us consider first
the anomalous magnetic dipole moment (MDM) of muon,
$a_\mu \equiv (g-2)/2$.
 The muon anomalous MDM
has been measured very precisely%
~\cite{Bennett:2006fi} as
\begin{eqnarray}
a_\mu^{\text{exp}} = 11659208.0(6.3)\times 10^{-10},
\end{eqnarray}
where the number in parentheses shows $1\sigma$ uncertainty.
 On the other hand,
the SM predicts
\begin{eqnarray}
a_\mu^{\text{SM}}[\tau] &=& 11659193.2(5.2)\times 10^{-10},\\
a_\mu^{\text{SM}}[e^+e^-] &=& 11659177.7(5.1)\times 10^{-10},
\end{eqnarray}
where the hadronic contributions
to $a_\mu^{\text{SM}}[\tau]$ and $a_\mu^{\text{SM}}[e^+e^-]$
were calculated~\cite{Davier:2009ag} by using data of
hadronic $\tau$ decay and $e^+e^-$ annihilation to hadrons,
respectively (See also
\cite{de Troconiz:2004tr,Hagiwara:2006jt,
DEHZ,Jegerlehner:2007xe,Prades:2009qp}).
 The deviations of the SM predictions from the experimental result
are given by
\begin{eqnarray}
\Delta a_\mu[\tau]
&\equiv& a_\mu^{\text{exp}} - a_\mu^{\text{SM}}[\tau]
 = 14.8(8.2)\times 10^{-10},\\
\Delta a_\mu[e^+e^-]
&\equiv& a_\mu^{\text{exp}} - a_\mu^{\text{SM}}[e^+e^-]
 = 30.3(8.1)\times 10^{-10}.
\end{eqnarray}
 These values of $\Delta a_\mu[\tau]$ and $\Delta a_\mu[e^+e^-]$
correspond to $1.8\sigma$ and $3.7\sigma$ deviations
from SM predictions, respectively.

 New contributions to $a_\mu$ at the 1-loop level in the HTM
come mainly from $H^{\pm\pm}$ and $H^{\pm}$.
 The Yukawa interactions of $H^\pm$,
which are mixtures of doublet and triplet Higgs bosons,
and of $H^{\pm\pm}$ ($= \Delta^{\pm\pm}$)
are written by
\begin{align}
{\mathcal L}_{\text{triplet-Yukawa}}^{H^\pm, H^{\pm\pm}}
= 
 -\sqrt{2} \frac{v}{\sqrt{v^2 + 2v_\Delta^2}}
  (U_\MNS^T h)_{i\ell}\, \overline{\nu_i^c} P_L \ell H^+
 -h_{\ell\ell^\prime} \overline{\ell^c} P_L \ell' H^{++}
+\text{H.c.},
\end{align}
where $P_L \equiv (1-\gamma^5)/2$ and
$\nu_i$ represent mass eigenstates of Majorana neutrinos
which satisfy conditions $\nu_i=\nu_i^c$.
 The 1-loop contribution of $H^\pm$
through the triplet Yukawa interaction%
\footnote
{
 The contribution through $m_\mu/v$
is ignored because it is suppressed by $v_\Delta^2/v^2$.
}
is calculated as
\begin{eqnarray}
a_\mu^{H^\pm}
&=&
 \frac{ m_\mu^2 }{ 8\pi^2 m_{H^\pm}^2 }
 \frac{v^2}{ v^2 + 2v_\Delta^2 }
 \sum_i
 (h^\dagger U_{\text{MNS}}^\ast)_{\mu i}
 (U_{\text{MNS}}^T h)_{i\mu}
\nonumber\\
&&\hspace*{40mm}
 \int_0^1\!\!\! dt\,
  \frac{ -t^2 (1-t) }
       {
        R_{H^\pm}^\mu t^2
        + ( 1 - R_{H^\pm}^\mu - R_{H^\pm}^i ) t
        + R_{H^\pm}^i
       }\\
&\simeq&
 -
 \frac{ \langle m^2 \rangle_{\mu\mu} }{ 96\pi^2 }
 \frac{ m_\mu^2 }{ v_\Delta^2 m_{H^\pm}^2 },
\end{eqnarray}
and the $H^{\pm\pm}$ contribution is given by
\begin{eqnarray}
a_\mu^{H^{\pm\pm}}
&=&
 \frac{ m_\mu^2 }{ 8\pi^2 m_{H^{\pm\pm}}^2 }
 \sum_\ell
 (h^\dagger)_{\mu\ell} h_{\ell\mu}
 \int_0^1\!\!\! dt
  \left[
   \frac{ -4t^2 (1-t) }
        {
         R_{H^{\pm\pm}}^\mu t^2
         + ( 1 - R_{H^{\pm\pm}}^\mu - R_{H^{\pm\pm}}^\ell ) t
         + R_{H^{\pm\pm}}^\ell
        }
  \right.
\nonumber\\
&&\hspace*{60mm}
  \left.
 {}+\frac{ -2t^2 (1-t) }
         {
          R_{H^{\pm\pm}}^\mu t^2
          + ( R_{H^{\pm\pm}}^\ell - R_{H^{\pm\pm}}^\mu - 1 ) t
          + 1
         }
  \right]\\
&\simeq&
 -
 \frac{ \langle m^2 \rangle_{\mu\mu} }{ 12\pi^2 }
 \frac{ m_\mu^2 }{ v_\Delta^2 m_{H^{\pm\pm}}^2 }.
\end{eqnarray}
 Here we have defined
\begin{eqnarray}
R^a_b
&\equiv&
\frac{m_a^2}{m_b^2},\\
\langle m^2 \rangle_{\ell\ell^\prime}
&\equiv&
 \left(
  U_{\text{MNS}}\,
  \text{diag}(m_1^2, m_2^2, m_3^2)\,
  U_{\text{MNS}}^\dagger
 \right)_{\ell\ell^\prime}
\ = \
 2 v_\Delta^2 (h^\dagger h)_{\ell\ell^\prime}.
\end{eqnarray}
 Note that $\langle m^2 \rangle_{\ell\ell^\prime}$ does not
depend on Majorana phases and
$\langle m^2 \rangle_{\ell\ell}$ is positive definite.
 Thus
$a_\mu^{\text{HTM}} \equiv a_\mu^{H^\pm} + a_\mu^{H^{\pm\pm}}$
is negative definite though $\Delta a_\mu$ is positive.
 The minus sign of the contributions
from Higgs triplets has been known~\cite{Abada:2007ux,tripletMDM}
but it does not seem to be taken seriously
probably because of confusions about
the combination $(h^\dagger h)_{\mu\mu}$.
 Usually the combination seems
to be written as $(h_{\mu\mu})^2$ or $(h^2)_{\mu\mu}$,
for which sign of $a_\mu^{\text{HTM}}$ can be flipped,
and then it seems possible to obtain
a (finite) constraint on $h_{\ell\ell^\prime}$.
 Actually,
any value of $h_{\ell\ell^\prime}$ can not fit
$\Delta a_\mu[e^+e^-]$ ($\Delta a_\mu[\tau]$)
at $3.7\sigma$ ($1.8\sigma$)
because of the wrong sign of $a_\mu^{\text{HTM}}$.

 Concerning only on the sign,
the 1-loop contribution from $H^0$ can have the right sign
to explain $\Delta a_\mu$
(see \cite{Dedes:2001nx} for the case in the type II
two Higgs doublet model (2HDM-II)).
 However,
the contribution has a suppression with $v_\Delta^2/v^2$ in the HTM
because $\Delta^0$ does not couple with charged leptons
at the tree level%
\footnote{
 The modification of the contribution
to $a_\mu^{\text{SM}}$ from $h^0$ is also suppressed
by $v_\Delta^2/v^2$.
}.
 Although in the Barr-Zee type~\cite{Barr-Zee} 2-loop diagrams
the right-sign contribution of $A^0$
can be important in some models like
the 2HDM-II~\cite{Cheung:2003pw}
and the minimal supersymmetric standard model (MSSM)%
~\cite{Cheung:2009fc},
such a situation does not happen in the HTM
because couplings of $A^0 (\simeq \text{Im}(\Delta^0))$
with quarks and charged leptons
are also suppressed by $v_\Delta/v$.

 The definite sign of $a_\mu^{\text{HTM}}$
is a feature of the simpleness and the predictability of the HTM\@.
 In the MSSM in contrast,
the contributions from supersymmetric particles to $a_\mu$
can have the right sign easily
by the appropriate choice of the sign
of the Higgs mass parameter $\mu_H$%
~\cite{Moroi:1995yh, Fukuyama:2003hn}.
 As the result,
the HTM is somehow disfavored by the muon anomalous MDM
and it results in a strong constraint on the model.
 This is also the case for other models
(e.g.\ the Zee-Babu model~\cite{Zee-Babu})
which do not have extra neutral Higgs bosons
with sizable couplings to charged leptons
similarly to the HTM\@.
 Of course,
the positive $\Delta a_\mu$ does not seem conclusive yet
and it does not mean exclusion of the HTM\@.
 The difference between $\Delta a_\mu[e^+e^-]$ and $\Delta a_\mu[\tau]$
may indicate existence of new physics in the quark sector
which is not modified in the HTM\@.

 Hereafter we take $m_{H^{\pm\pm}} = m_{H^\pm}$ for simplicity.
 The large splitting of their masses is disfavored
by the constraint on the $\rho$ parameter.
 Once we fix the neutrino mass matrix,
muon anomalous MDM and LFV processes
are interpreted as lower bounds on $v_\Delta m_{H^{\pm\pm}}$.
\begin{figure}[t]
\begin{center}
\includegraphics[origin=c, angle=-90, scale=0.3]
{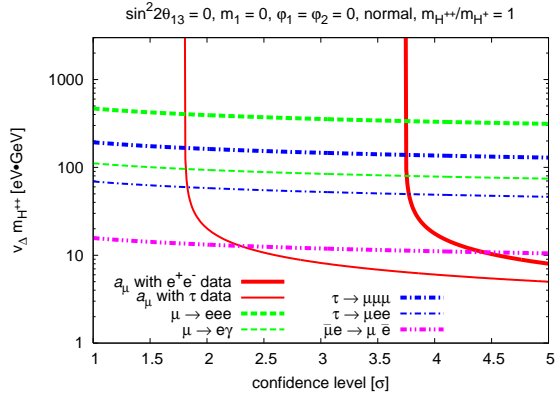}
\vspace*{-10mm}
\caption{
 Lower bounds on $v_\Delta m_{H^{\pm\pm}}$
given by constraints on muon anomalous MDM and LFV processes
as functions of the confidence level.
 All parameters in the neutrino mass matrix are fixed
as an example (see the text for the values)
for the normal mass ordering.
 We take $m_{H^{\pm\pm}} = m_{H^\pm}$ for simplicity.
}
\label{fig:CL}
\end{center}
\end{figure}
 Figure~\ref{fig:CL} shows the lower bounds with respect to
the confidence level in a unit of the standard deviation $\sigma$
and they are given by constraints on
the muon anomalous MDM with $e^+e^-$ data (bold solid red line),
the MDM with $\tau$ data (solid red line),
$\mu \to \bar{e} e e$ (bold dashed green line),
$\mu \to e \gamma$ (dashed green line),
$\tau \to \bar{\mu} \mu \mu$ (bold dash-dotted blue line),
$\tau \to \bar{\mu} e e$ (dash-dotted blue line),
and the muonium ($\mu^+ e^-$) conversion
to the anti-muonium (bold dash-dot-dotted magenta line).
 Bounds from $\tau \to \bar{\mu} \mu \mu$
and $\tau \to \bar{\mu} e e$ are important in our analysis
among six possible $\tau \to \bar{\ell}\ell^\prime \ell^{\prime\prime}$.
 Formulae of branching ratios of these LFV decays in the HTM
and their current bounds at 90\%~CL are
\begin{eqnarray}
\BR(\mu \to \bar{e} e e)
&=&
 \frac{ |m_{\mu e}|^2 |m_{ee}|^2 }
      { 16 G_F^2 v_\Delta^4 m_{H^{\pm\pm}}^4 }
 < 1.0\times 10^{-12} \
\text{\cite{Bellgardt:1987du}},
\label{bound:meee}\\
\BR(\mu \to e \gamma)
&=&
 \frac{ 27 \alpha | \langle m^2 \rangle_{e\mu}|^2 }
      { 256 \pi G_F^2 v_\Delta^4 m_{H^{\pm\pm}}^4 }
 < 1.2\times 10^{-11} \
\text{\cite{Brooks:1999pu}},
\label{bound:meg}\\
\BR(\tau \to \bar{\mu} \mu \mu)
&=&
 \frac{ |m_{\tau\mu}|^2 |m_{\mu\mu}|^2 }
      { 16 G_F^2 v_\Delta^4 m_{H^{\pm\pm}}^4 }\,
 \BR(\tau\to \mu\bar{\nu}_\mu\nu_\tau)
 < 3.2\times 10^{-8} \
\text{\cite{Miyazaki:2007zw}},
\label{bound:tmmm}\\
\BR(\tau \to \bar{\mu} ee)
&=&
 \frac{ |m_{\tau\mu}|^2 |m_{ee}|^2 }
      { 16 G_F^2 v_\Delta^4 m_{H^{\pm\pm}}^4 }\,
 \BR(\tau\to \mu\bar{\nu}_\mu\nu_\tau)
 < 2.0\times 10^{-8} \
\text{\cite{Miyazaki:2007zw}},
\label{bound:tmee}
\end{eqnarray}
where $\BR(\tau\to \mu\bar{\nu}_\mu\nu_\tau) = 17\%$,
$\alpha = 1/137$ stands for the fine structure constant,
and $G_F = 1.17\times 10^{-5}\,\GeV^{-2}$ denotes
the Fermi coupling constant.
 The effective Lagrangian for
the muonium conversion is
\begin{eqnarray}
{\cal L}_{M\overline{M}}
\ = \
 2\sqrt{2} G_{M\overline{M}}
 \left( \overline{\mu} \gamma^\rho P_L e \right)
 \left( \overline{\mu} \gamma_\rho P_L e \right)
\ = \
 4\sqrt{2} G_{M\overline{M}}
 \left( \overline{\mu} P_R \mu^c \right)
 \left( \overline{e^c} P_L e \right).
\end{eqnarray}
 The formula of the coupling $G_{M\overline{M}}$ in the HTM
and current constraint at 90\%~CL for that are
\begin{eqnarray}
\left( \frac{|G_{M\bar{M}}|}{G_F} \right)^2
&=&
 \frac{ |m_{ee}|^2 |m_{\mu\mu}|^2 }
      { 128 G_F^2 v_\Delta^4 m_{H^{\pm\pm}}^4 }
 < ( 3.0\times 10^{-3} )^2 \
\text{\cite{Willmann:1998gd}}.
\label{bound:muonium} 
\end{eqnarray}

 In Fig.~\ref{fig:CL},
parameters of the neutrino mass matrix
are fixed by (\ref{param1}), (\ref{param2}),
and the following values as an example:
$m_1=0$,
$\sin^2{2\theta_{13}} = 0$,
$\varphi_1 = \varphi_2 = 0$.
 With these values of parameters,
we have
$\langle m^2 \rangle_{\mu\mu} = 1.2\times 10^{-3}\,\eV^2$,
$|m_{\mu e}|^2 |m_{ee}|^2 = 6.4\times 10^{-11}\,\eV^4$,
$|\langle m^2 \rangle_{e\mu}|^2 = 6.3\times 10^{-10}\,\eV^4$,
$|m_{\tau\mu}|^2 |m_{\mu\mu}|^2 = 3.5\times 10^{-7}\,\eV^4$,
$|m_{\tau\mu}|^2 |m_{ee}|^2 = 3.6\times 10^{-9}\,\eV^4$,
and $|m_{ee}|^2 |m_{\mu\mu}|^2 = 5.9\times 10^{-9}\,\eV^4$.
 Bounds (\ref{bound:meee})-(\ref{bound:tmee})
and (\ref{bound:muonium}) at 90\%~CL
are translated naively
into $x \sigma$~CL bounds by multiplying $x/1.64$
because 90\%~CL corresponds to $1.64\sigma$.
 Below around $1.8\sigma$ ($3.7\sigma$),
the muon anomalous MDM $\Delta a_\mu[\tau]$ ($\Delta a_\mu[e^+e^-]$)
gives the strongest constraint on the HTM
but it becomes weaker rapidly than other constraints
at higher confidence levels.
 Hereafter,
we take $\Delta a_\mu[\tau]$ and concentrate ourselves
on $2\sigma$~CL in order to avoid qualitative
disagreement with $\Delta a_\mu[\tau]$ in the HTM\@.

\begin{figure}[t]
\begin{center}
\includegraphics[origin=c, angle=-90, scale=0.3]
{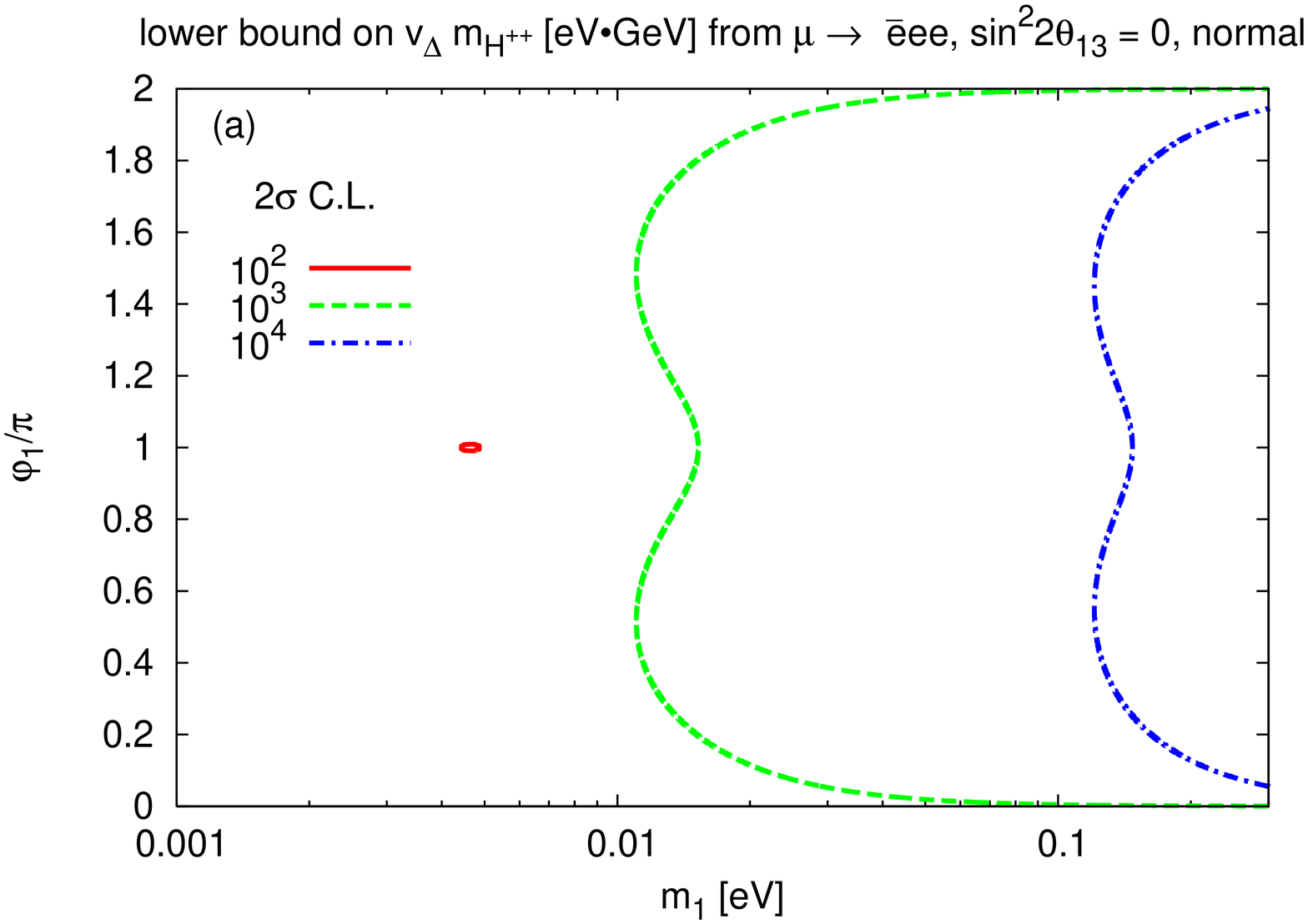}
\includegraphics[origin=c, angle=-90, scale=0.3]
{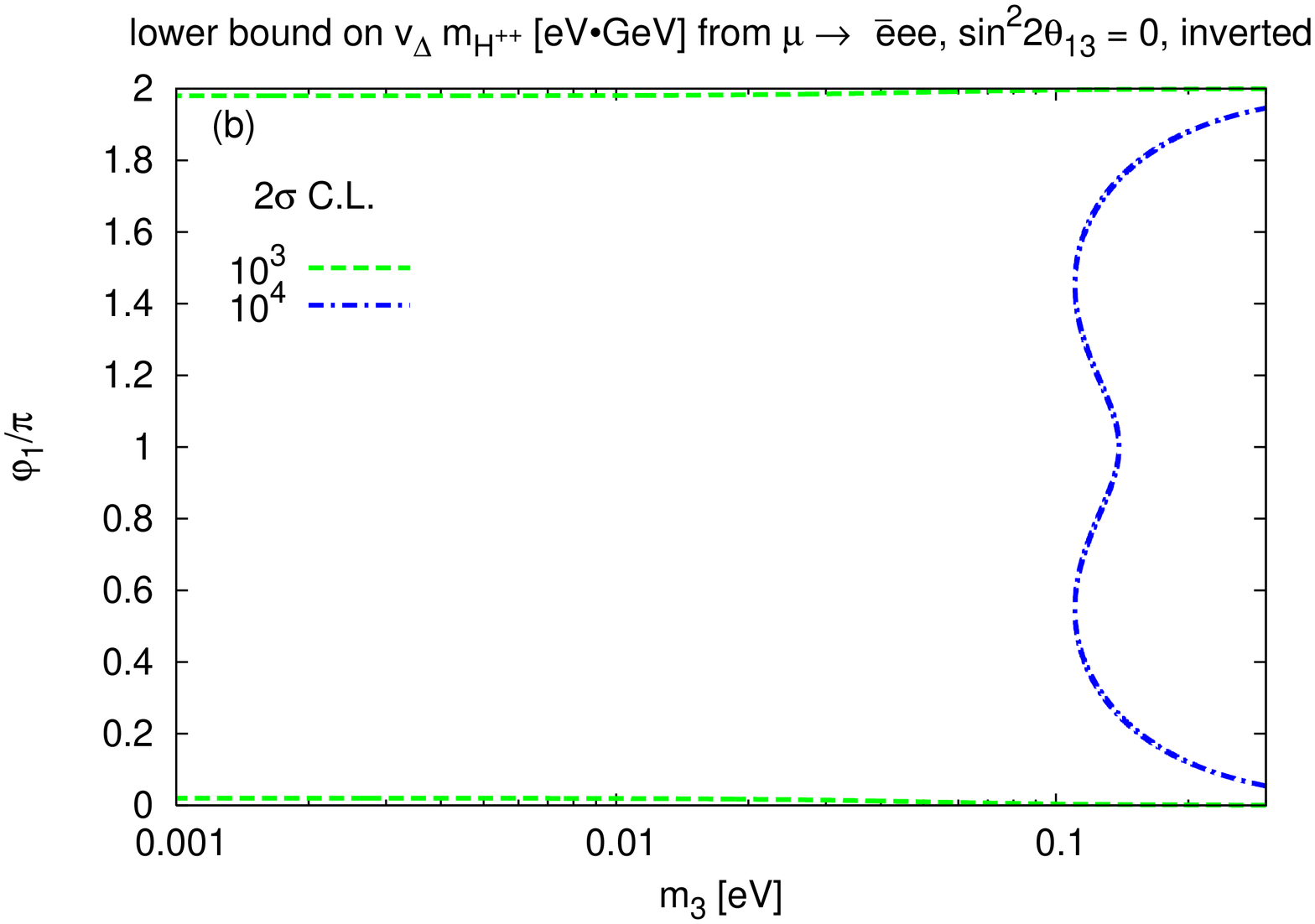}
\vspace*{-10mm}
\caption{Contours of the lower bounds on $\vm [\eVGeV]$
given by $\meee$.
(a) for the normal mass ordering.
(b) for the inverted mass ordering. }
\label{fig:meee}
\end{center}
\end{figure}
%
\begin{figure}[t]
\begin{center}
\includegraphics[origin=c, angle=-90, scale=0.3]
{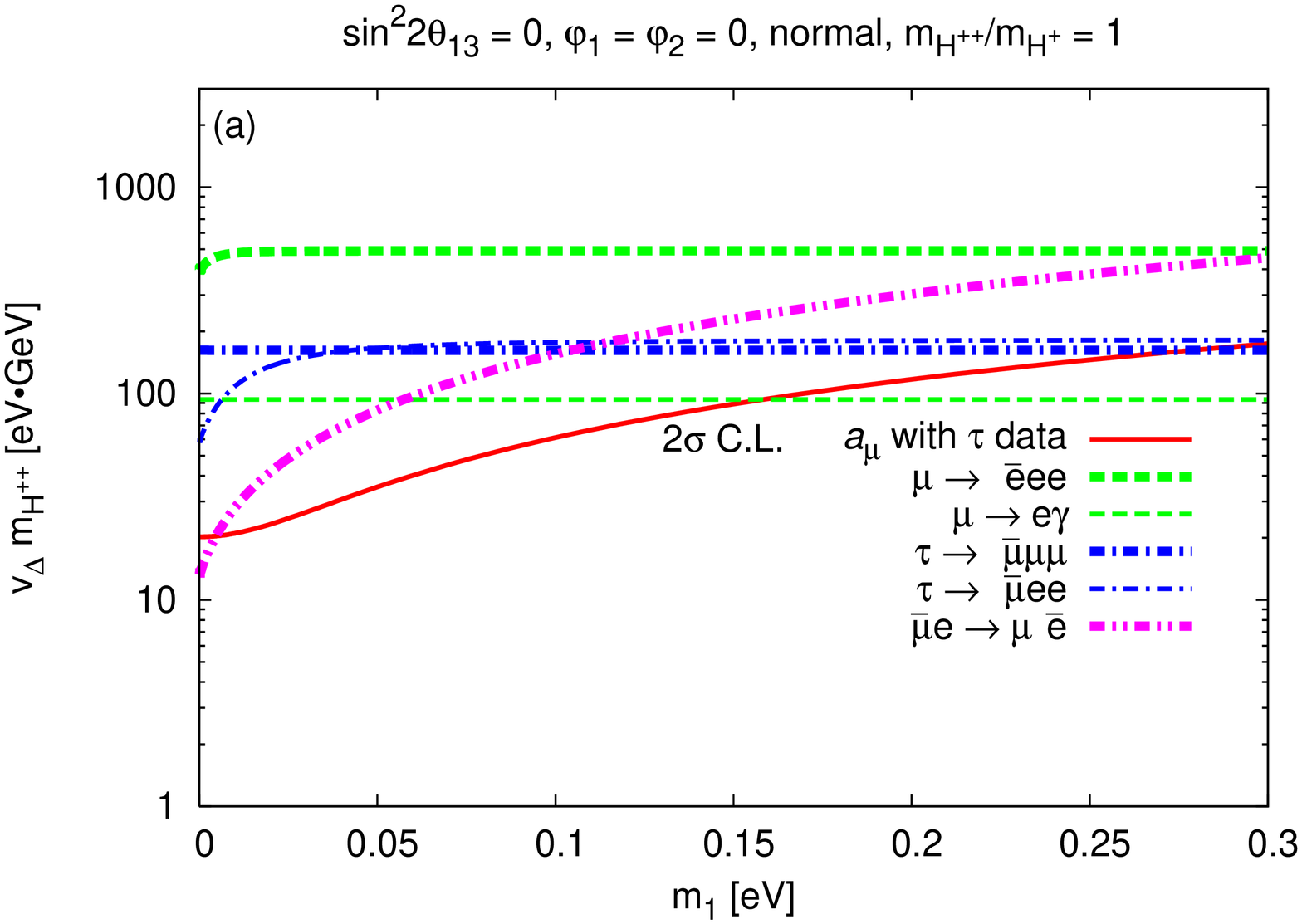}
\includegraphics[origin=c, angle=-90, scale=0.3]
{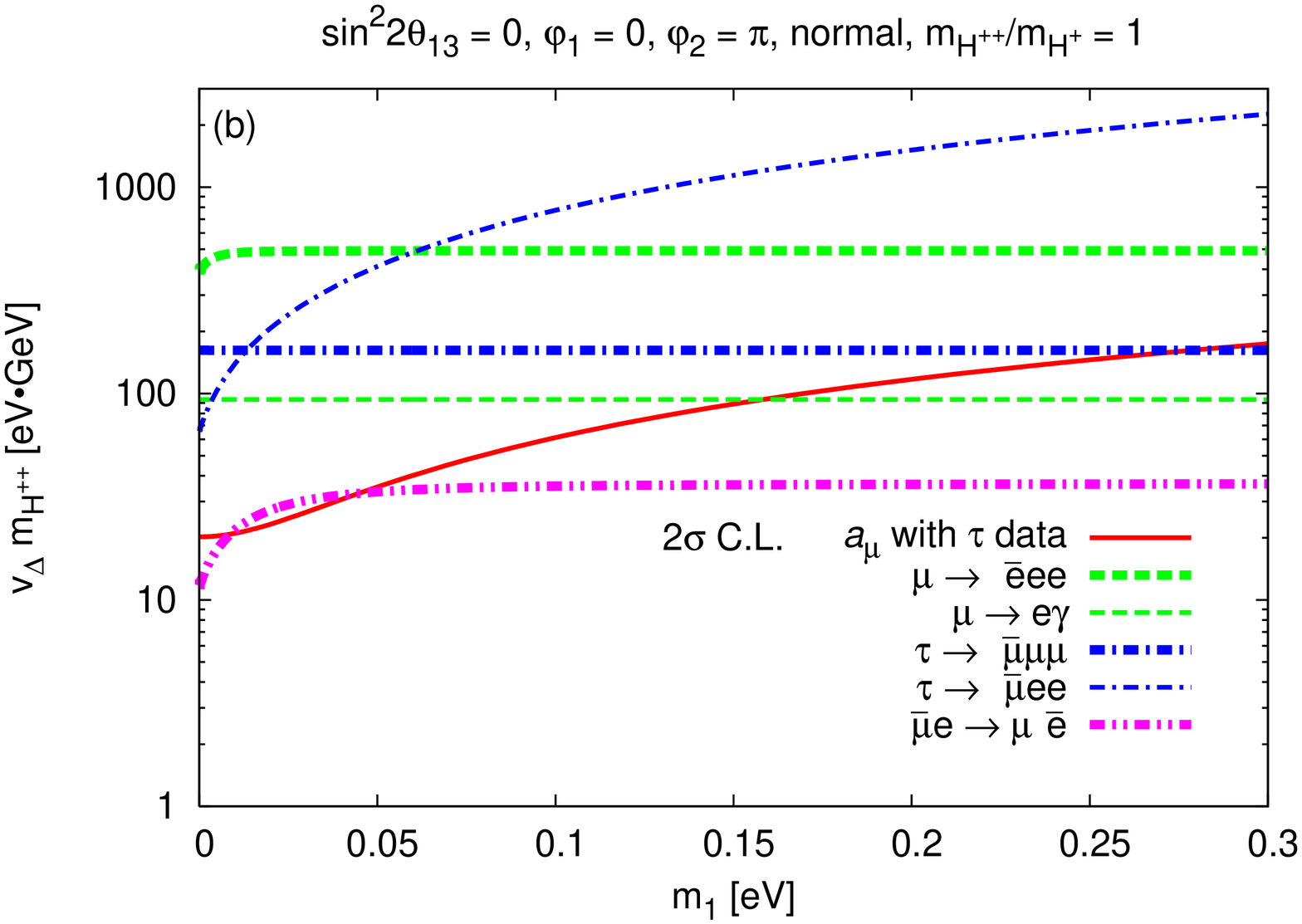}
\vspace*{-10mm}
\caption{Lower bounds on $\vm$ for $\varphi_1 = 0$
in the normal mass ordering.
(a) for $\varphi_2 = 0$.
(b) for $\varphi_2 = \pi$.}
\label{fig:2sigma_n}
\end{center}
\end{figure}
%
\begin{figure}[t]
\begin{center}
\includegraphics[origin=c, angle=-90, scale=0.3]
{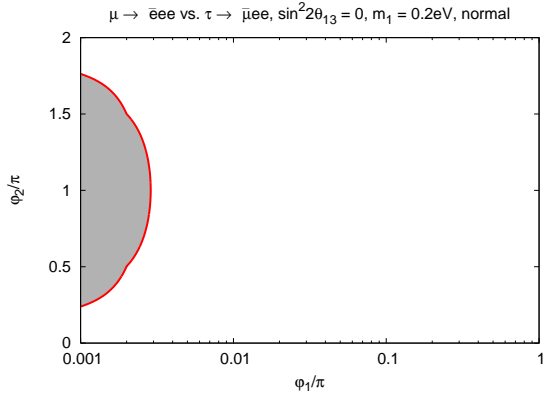}
\vspace*{-10mm}
\caption{In the shaded region,
the bound on $\vm$ from $\tmee$ is stronger than
the one from $\meee$.
 We used $m_1 = 0.2\,\eV$ for the normal mass ordering.
The shaded region is symmetric under
a transformation of
$(\varphi_1, \varphi_2-\pi ) \to ( -\varphi_1, -\varphi_2 + \pi )$.}
\label{fig:vmrat_n}
\end{center}
\end{figure}

\subsection{Constraints in the case of $\BR(\meee) \neq 0$}

 In most of parameter space,
the strongest lower bound on $\vm$ is given by
$\meee$ as expected naively from the strong constraint
on its branching ratio (\ref{bound:meee}).
 Figures~\ref{fig:meee}(a) and (b) show contours of the bounds
with $\theta_{13}=0$
for the normal and inverted mass ordering, respectively.
 Note that $\BR(\meee)$ does not depend on
$\delta$ and $\varphi_2$ for $\theta_{13} = 0$.
 Although the bound on $\vm$ from $\meee$
is relatively weak for small $m_1$ in Fig.~\ref{fig:meee}(a),
bounds from other LFV processes are weaker than that.
 It is shown also that $\varphi_1 \simeq 0$
makes the bound from $\meee$ weak for both of mass orderings.
 We focus on the case of $\varphi_1 = 0$
in the next paragraph.
 In Fig.~\ref{fig:meee}(a)
there is a special point at $\varphi_1 = \pi$ and
$m_1 = s_{12}^2 \sqrt{\Delta m^2_{21}}/\sqrt{\cos{2\theta_{12}}}
\simeq 4.6\times 10^{-3}\,\eV$
where the bound vanishes because of $m_{ee} = 0$.
 Such cases of $\BR (\meee) = 0$ are discussed
in the next subsection.

 In Fig.~\ref{fig:2sigma_n},
$m_1$-dependences of bounds
from $\Delta a_\mu[\tau]$ and LFV processes
are presented for the normal mass ordering
at $\varphi_1 = 0$
where the bound from $\meee$ is relatively weak.
 Figures~\ref{fig:2sigma_n}(a) and (b)
are obtained for $\varphi_2 = 0$ and $\pi$,
respectively.
 Other parameters are the same values
as ones in Fig.~\ref{fig:CL}.
 It is seen in Fig.~\ref{fig:2sigma_n}(a) that
$\meee$ still gives the most stringent bound
for $\varphi_1 = \varphi_2 = 0$ and $m_1 \lesssim 0.3\,\eV$
although the bound from the muonium conversion
gets close to that for large $m_1$.
 If we accept $m_1\gtrsim 0.3\,\eV$,
the bound from the muonium conversion
can be stronger than the bound from $\meee$.
 On the other hand,
Fig.~\ref{fig:2sigma_n}(b) shows that
the bound from $\tmee$ can be more stringent than
the one from $\meee$ for $m_1 \gtrsim 0.06\,\eV$.
 This is because a parameter set
$(\theta_{13}, \varphi_1, \varphi_2) = (0, 0, \pi)$
in the region of $\Delta m^2_{ij}/m_1^2 \ll 1$
gives
\begin{eqnarray}
|m_{\mu e}|^2 |m_{ee}|^2
&\simeq&
 \frac{1}{32} (\Delta m^2_{21})^2 \sin^2{2\theta_{12}}
\simeq 1.6\times 10^{-10}\,\eV^4,
\label{eq:meee_vs}\\
|m_{\tau\mu}|^2 |m_{ee}|^2
&\simeq&
 m_1^4,
\label{eq:tmee_vs}
\end{eqnarray}
and the large difference between
experimental constraints (\ref{bound:meee}) and (\ref{bound:tmee})
can be compensated for $m_1 \gtrsim O(0.1)\,\eV$.
 In Fig.~\ref{fig:vmrat_n}
the shaded region shows
values of Majorana phases
for which the bound from $\tmee$ becomes
more stringent than the one from $\meee$
at $m_1 = 0.2\,\eV$ for the normal mass ordering.
 The region is symmetric under a transformation of
$( \varphi_1, \varphi_2 - \pi ) \to ( -\varphi_1, - \varphi_2 + \pi )$
because of $|m_{\ell\ell^\prime}| = |m_{\ell\ell^\prime}^\ast|$.
 Although the bound from $\meee$ is relatively
weak for $\varphi_1 \simeq 0$,
the bound is still the most stringent one
at around $\varphi_2 = 0$ because
$\tmee$ is also suppressed.
 If we take nonzero $\theta_{13}$ and ignore $\Delta m^2_{21}$
for $(\varphi_1, \varphi_2) = (0, \pi)$,
eq.~(\ref{eq:meee_vs}) is rewritten as
\begin{eqnarray}
 |m_{\mu e}|^2 |m_{ee}|^2
&\simeq&
 2 s_{13}^2 m_1^4
\end{eqnarray}
while eq.~(\ref{eq:tmee_vs}) remains valid.
 Therefore,
the shaded region in Fig.~\ref{fig:vmrat_n}
at around $(\varphi_1, \varphi_2) = (0, \pi)$
exists for $\sin^2{2\theta_{13}} \lesssim 10^{-5}$.
 For the inverted mass ordering,
the region where $\tmee$ becomes remarkable
is almost same as the one
in Figs.~\ref{fig:2sigma_n} and \ref{fig:vmrat_n}
because neutrino masses are almost degenerated in the region.
 In such a region,
we can also expect a signal of $\tmee$ in future experiments%
~\cite{superB, LHCb}
with satisfying the current constraint on $\meee$.

\subsection{Constraints in Cases of $\BR(\meee) = 0$}
 It has been known that
the strong constraint from $\meee$ can be evaded in the cases of
$m_{e\mu} = 0$~\cite{Chun:2003ej}
and $m_{ee} =  0$~\cite{Akeroyd:2009nu}.
 While it is impossible to have $m_{e\mu} = 0$
with $\theta_{13} = 0$,
the case of $m_{ee} = 0$ is possible also for $\theta_{13} = 0$
as we mentioned for Fig.~\ref{fig:meee}(a).
 Such cancellations in the HTM are desired
also for experiments~\cite{superB,LHCb,Adam:2009ci}
to discover some LFV decays ($\mu \to e\gamma$ etc.)%
~\cite{Chun:2003ej, Akeroyd:2009nu} in the future.
\begin{figure}[t]
\begin{center}
\includegraphics[origin=c, angle=-90, scale=0.3]
{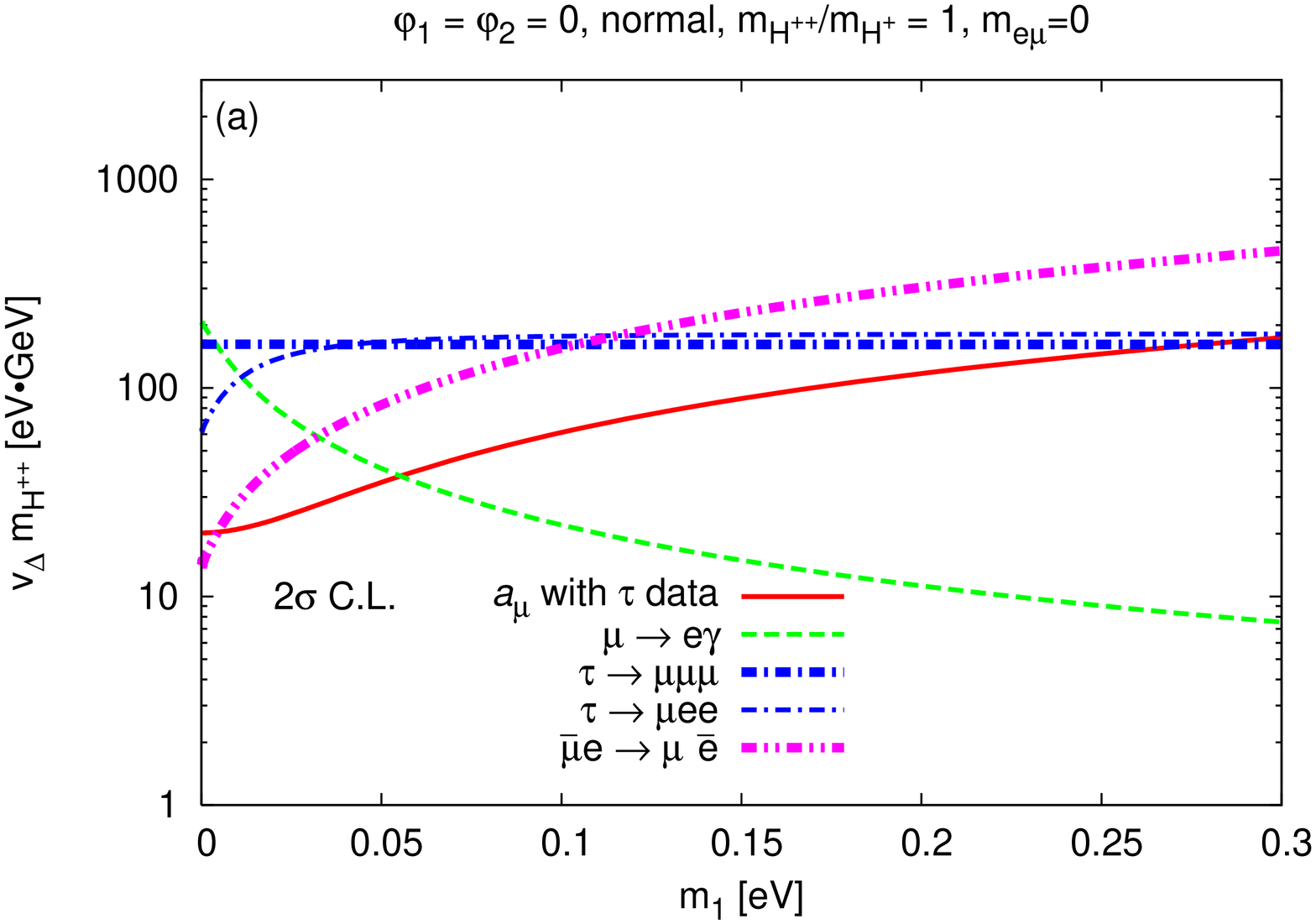}
\includegraphics[origin=c, angle=-90, scale=0.3]
{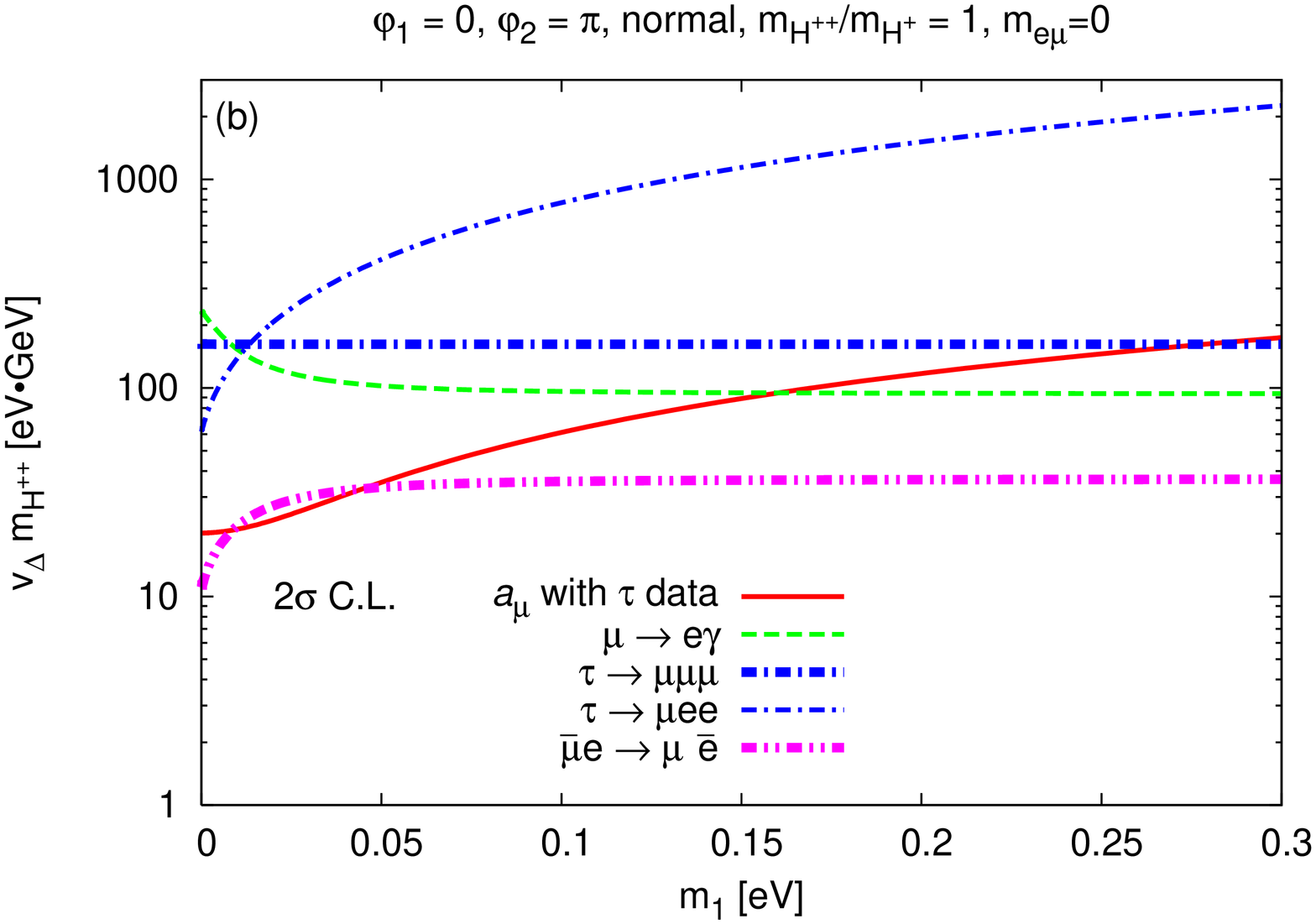}\\[5mm]
\includegraphics[origin=c, angle=-90, scale=0.3]
{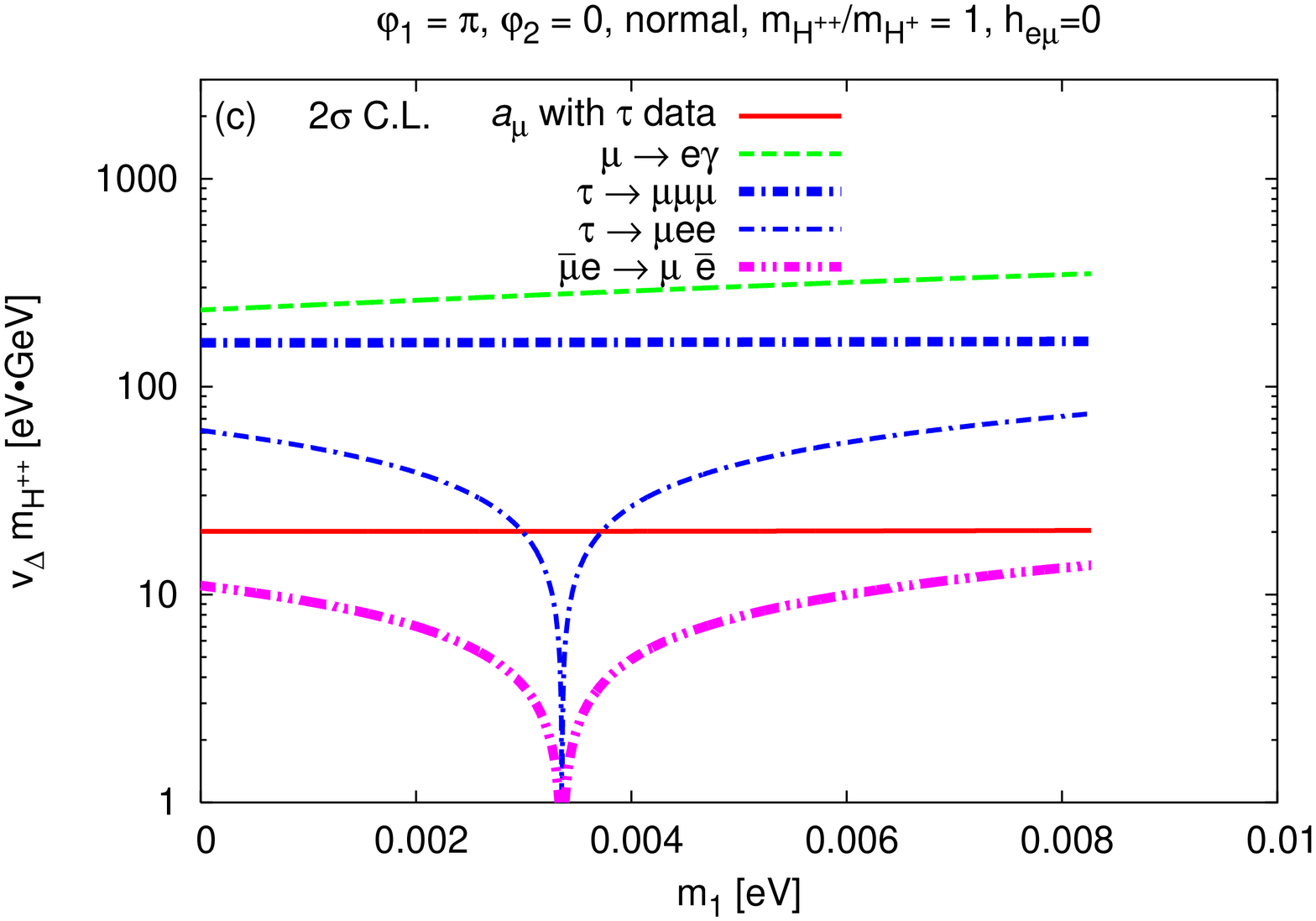}
\includegraphics[origin=c, angle=-90, scale=0.3]
{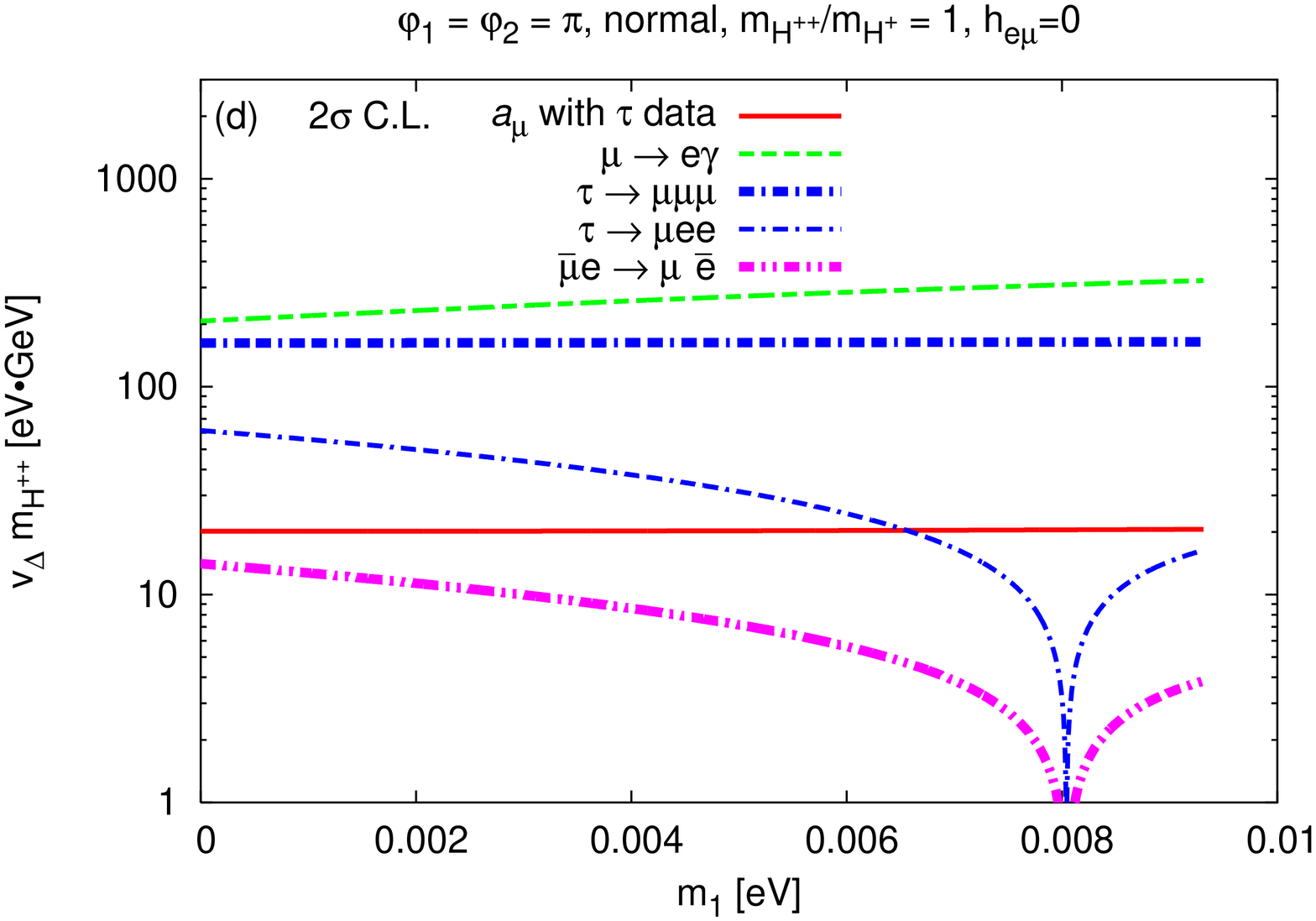}
\vspace*{-10mm}
\caption{ Lower bounds on $\vm$
for cases of $m_{e\mu}=0$ in the normal mass ordering.
The value of $\theta_{13}^\magic$ varies to keep $m_{e\mu}=0$
(See Appendix).
(a) for $(\varphi_1, \varphi_2) = (0, 0), \delta^\magic = \pi$.
(b) for $(\varphi_1, \varphi_2) = (0, \pi), \delta^\magic = 0$.
(c) for $(\varphi_1, \varphi_2) = (\pi, 0), \delta^\magic = 0$.
(d) for $(\varphi_1, \varphi_2) = (\pi, \pi), \delta^\magic = \pi$.
}
\label{fig:hem0_n}
\end{center}
\end{figure}
 Figures~\ref{fig:hem0_n}(a)-(d) show results
for the case of $m_{e\mu}=0$ in the normal mass ordering.
 Four CP conserving sets of Majorana phases
are taken for the figures as examples.
 We use appropriate values of $\theta_{13}$ and $\delta$
for $m_{e\mu}=0$,
which we call as ``magic values''
$\theta_{13}^\magic$ and $\delta^\magic$,
and explicit formulae of them are shown in Appendix.
 For each cases in Fig.~\ref{fig:hem0_n},
the magic value $\delta^\magic$ is $0$ or $\pi$
independently of $m_1$.
 Although $\langle m^2 \rangle_{e\mu}$ is independent
of $m_1$ and Majorana phases,
the bounds from $\mu \to e\gamma$ in Fig.~\ref{fig:hem0_n}
are not constant with respect to these parameters
because $\theta_{13}^\magic$ depends on them.
 We see in Figs.~\ref{fig:hem0_n}(a)-(d)
that the bound from $\meg$ is the strongest one
for $m_1\simeq 0$,
and this is also the case with any values
of $\varphi_1$ and $\varphi_2$.
 For the case of Fig~\ref{fig:hem0_n}(a),
$\tau$ decays give the most stringent bound
for $0.01\,\eV \lesssim m_1 \lesssim 0.12\,\eV$,
and the bound from the muonium conversion
becomes the strongest one for $m_1 \gtrsim 0.12\,\eV$.
 In Fig.~\ref{fig:hem0_n}(b),
the bound from $\tau \to \bar{\mu}ee$ is prominent.
 The magic $\theta_{13}$ in Fig.~\ref{fig:hem0_n}(b)
gives $\sin^2{2\theta_{13}^\magic} \simeq 10^{-7}$
for $m_1=0.2\,\eV$,
and then the remarkable behavior of the bound
from $\tmee$ can be understood also as
a part of the case shown in Fig.~\ref{fig:vmrat_n}
whose shaded region appears for
$\sin^2{2\theta_{13}} \lesssim 10^{-5}$.
 In Figs.~\ref{fig:hem0_n}(c) and (d),
the most stringent bound is obtained from $\mu \to e\gamma$.
 Note that $\sin^2{2\theta_{13}^\magic}$
in Figs.~\ref{fig:hem0_n}(c) and (d)
become larger than 0.14 of the CHOOZ bound%
~\cite{Apollonio:2002gd}
for $m_1 \gtrsim 0.008\,\eV$,
and then we can not have $m_{e\mu} = 0$ for the case.

\begin{figure}[t]
\begin{center}
\includegraphics[origin=c, angle=-90, scale=0.3]
{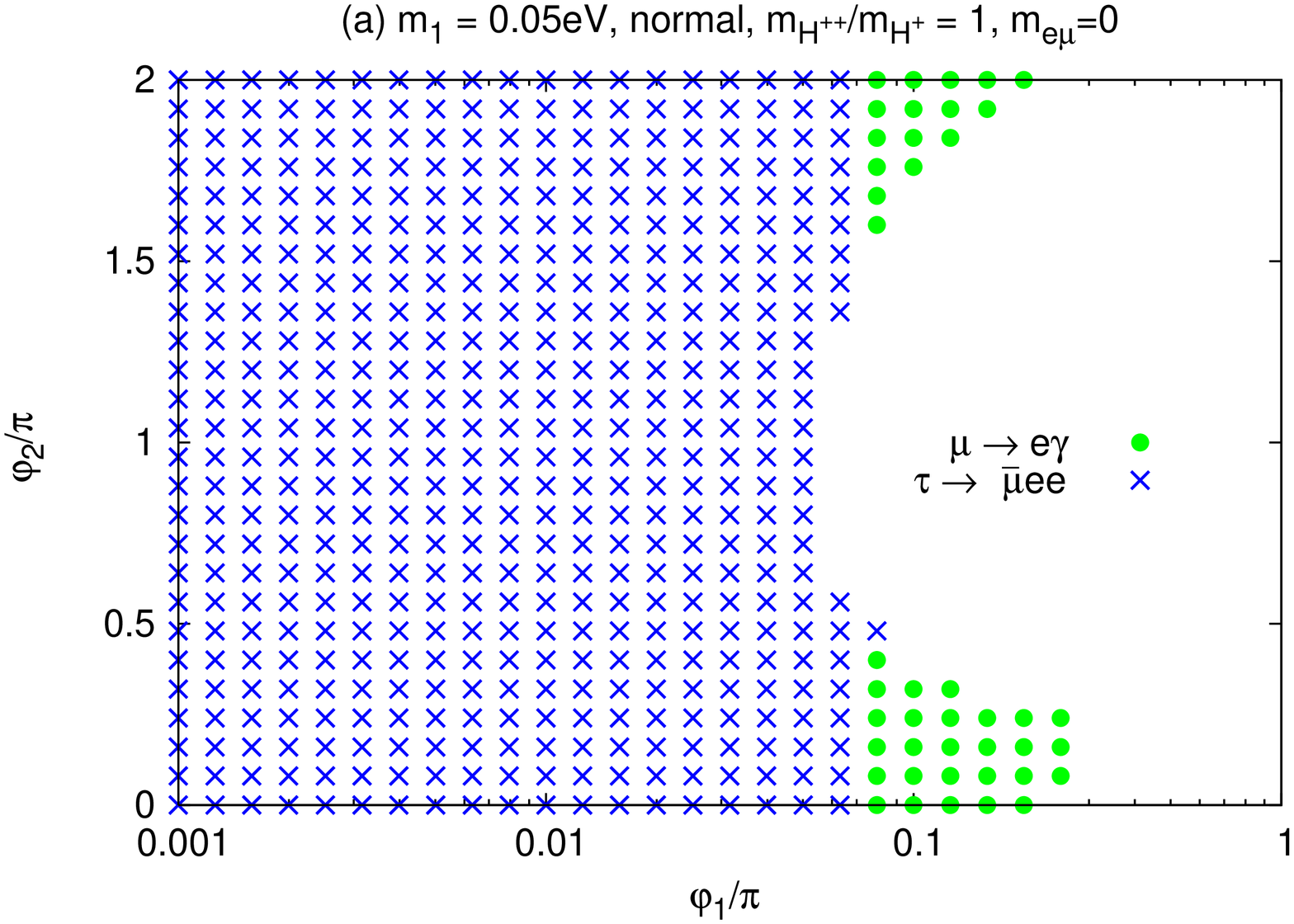}
\includegraphics[origin=c, angle=-90, scale=0.3]
{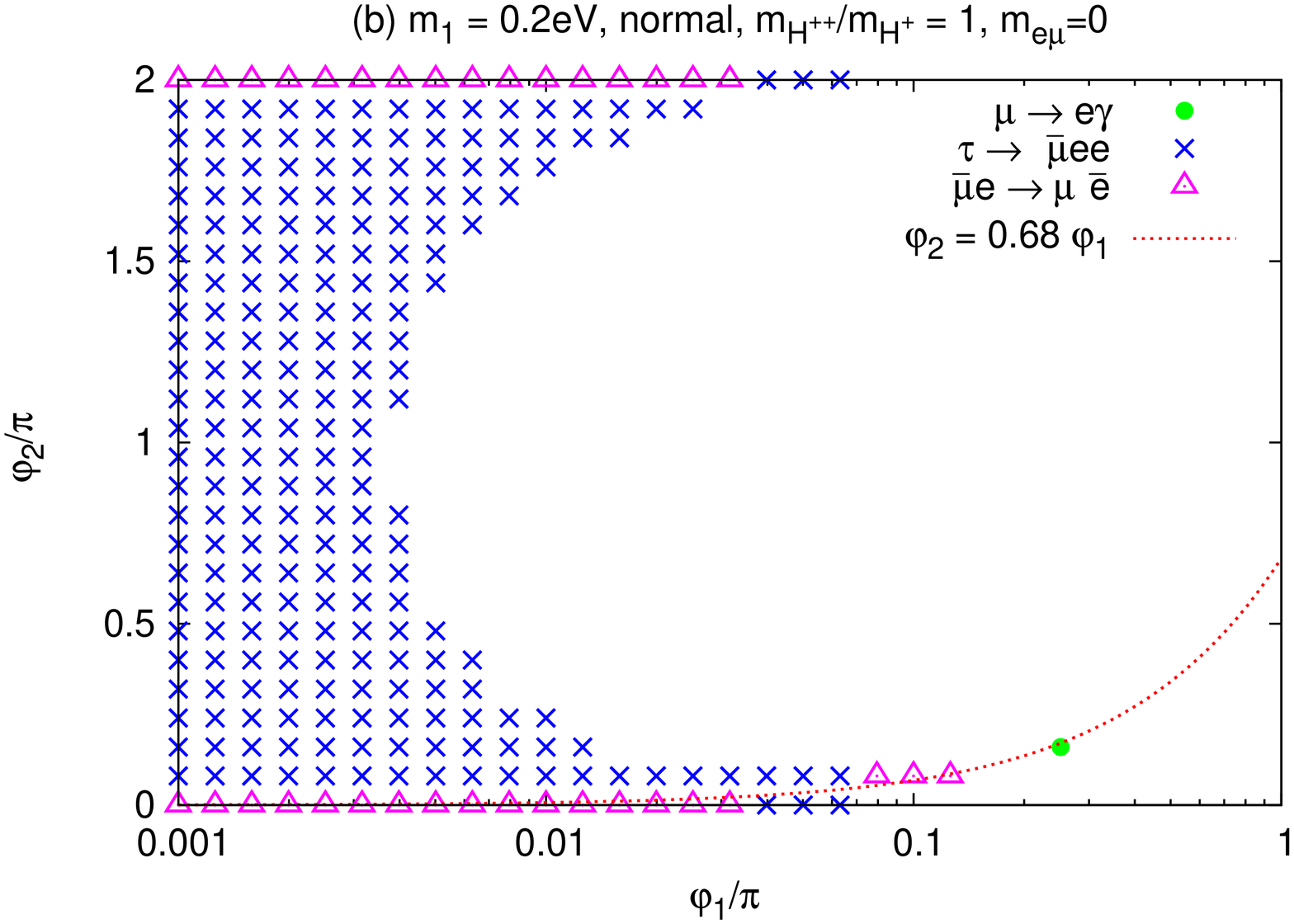}
\vspace*{-10mm}
\caption{
 Green circles, blue crosses, and magenta triangles
show regions where the strongest lower bound on $\vm$
for the case of $m_{e\mu} = 0$ in the normal mass ordering
comes from $\meg$, $\tmee$, and the muonium conversion,
respectively.
 The values of $\theta_{13}^\magic$ and $\delta^\magic$ are
functions of Majorana phases (See Appendix).
 We can not have $m_{e\mu}=0$ outside of these regions.
 The regions are symmetric under a transformation
$(\varphi_1, \varphi_2) \to (-\varphi_1, -\varphi_2)$.
(a) with $m_1=0.05\,\eV$.
(b) with $m_1 = 0.2\,\eV$.
 With $m_1 = 0$,
the most stringent bound is given by $\meg$
for all values of $\varphi_1$ and $\varphi_2$.
}
\label{fig:mphase_hem0_n}
\end{center}
\end{figure}
 Figure~\ref{fig:mphase_hem0_n} shows
which process gives the most stringent lower bound on $\vm$
in a space of Majorana phases
by keeping $m_{e\mu}=0$ for the normal mass ordering.
 Green circles, blue crosses, and magenta triangles
show regions where the strongest bound
comes from $\meg$, $\tmee$, and the muonium conversion,
respectively.
 It is impossible to achieve $m_{e\mu}=0$
outside of the regions because of
unacceptably large $\theta_{13}^\magic$,
and then it becomes the situations
discussed in the previous subsection
for $\BR(\meee)\neq 0$.
 Figure~\ref{fig:mphase_hem0_n}(a) is for $m_1=0.05\,\eV$
and (b) is for $m_1=0.2\,\eV$.
 For $m_1=0$,
it is possible to have $m_{e\mu}=0$
for any values of Majorana phases,
and the most stringent bound is always given by $\meg$
as $\vm \gtrsim 200\,\eVGeV$.
 Values of lower bounds on $\vm$ for Fig.~\ref{fig:mphase_hem0_n}(a)
from  $\meg$ and $\tmee$
vary in $150\to 350\,\eVGeV$ and $150\to 400\,\eVGeV$, respectively.
 We have
$\vm \gtrsim 300\,\eVGeV$, $\gtrsim 300\to 1500\,\eVGeV$,
and $\gtrsim 300\,\eVGeV$ in Fig.~\ref{fig:mphase_hem0_n}(b)
from  $\meg$, $\tmee$, and the muonium conversion,
respectively.
 The regions are symmetric under a transformation
$(\varphi_1, \varphi_2) \to (-\varphi_1, -\varphi_2)$
because of $|m_{\ell\ell^\prime}| = |m_{\ell\ell^\prime}^\ast|$.
 We see that $\varphi_1 \simeq 0$
is preferred to keep $s_{13}^\magic$ small for $m_1\neq 0$
and we can confirm $s_{13}^\magic \propto \Delta m^2_{21}$
for $\varphi_1 = 0$ with eq.~(\ref{eq:s13mgc}) in Appendix.
 It can be found also with eq.~(\ref{eq:s13mgc})
for $\varphi_1, \varphi_2 \ll 1$ and $\Delta m^2_{ij} = 0$
that
\begin{eqnarray}
\varphi_2
 \simeq \frac{1 + \cos{2\theta_{12}}}{2}\,\varphi_1
 = 0.68\,\varphi_1 \ \ \
(\text{red dotted line in Fig.~\ref{fig:mphase_hem0_n}(b)})
\end{eqnarray}
is preferred to have a small $s_{13}^\magic$.
 Bounds from $\tmee$ and the muonium conversion
can be the most stringent one
only for $\varphi_1 \lesssim 0.1\pi$.
 Majorana phases are almost restricted
as $\varphi_2 \simeq 0.68\,\varphi_1$
for the case of a strong constraint from
the muonium conversion.
 It is shown that $\meg$ can be the most stringent bound
even for $m_1 \gtrsim 0.008\,\eV$ (cf.\ Fig.~\ref{fig:hem0_n})
because of large $s_{13}^\magic$.
 At the border to the white region in Fig.~\ref{fig:mphase_hem0_n},
we have $\sin^2{2\theta_{13}^\magic} = 0.14$.

\begin{figure}[t]
\begin{center}
\includegraphics[origin=c, angle=-90, scale=0.3]
{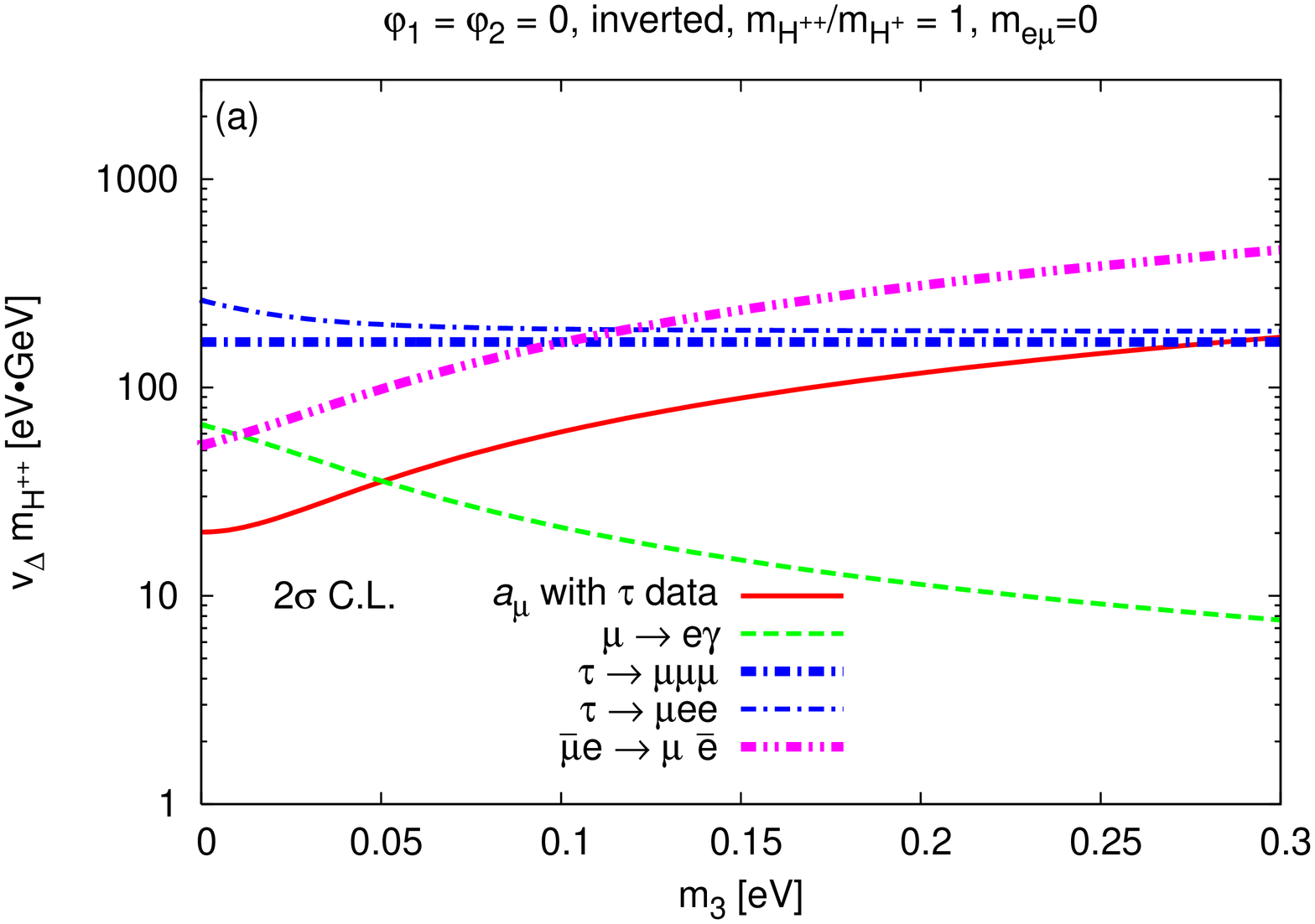}
\includegraphics[origin=c, angle=-90, scale=0.3]
{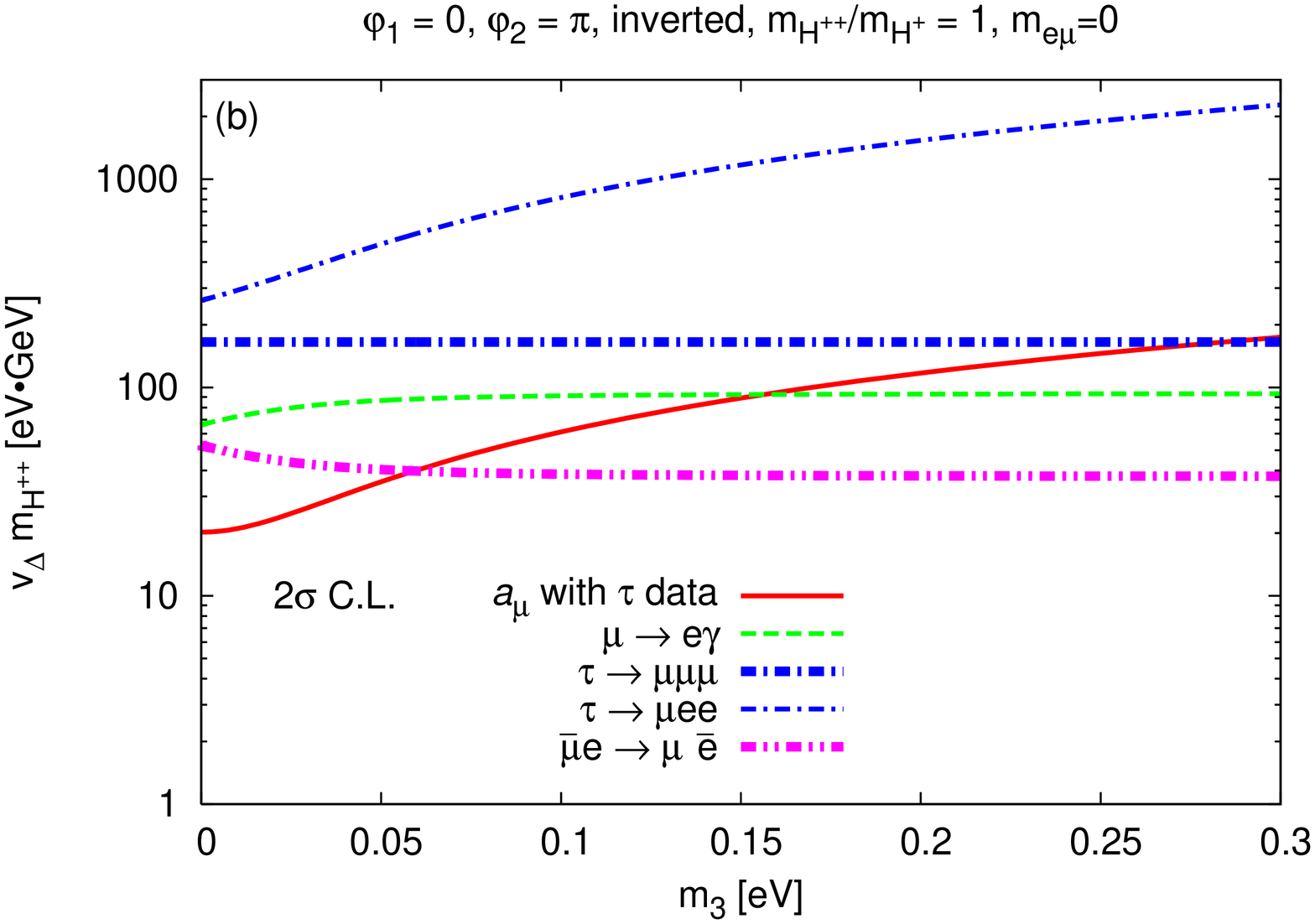}
\vspace*{-10mm}
\caption{ Lower bounds on $\vm$
for cases of $m_{e\mu}=0$ in the inverted mass ordering.
The value of $\theta_{13}^\magic$ varies to keep $m_{e\mu}=0$
(See Appendix).
(a) for $(\varphi_1, \varphi_2) = (0, 0), \delta^\magic = 0$.
(b) for $(\varphi_1, \varphi_2) = (0, \pi), \delta^\magic = 0$.
}
\label{fig:hem0_i}
\end{center}
\end{figure}
%
\begin{figure}[t]
\begin{center}
\includegraphics[origin=c, angle=-90, scale=0.3]
{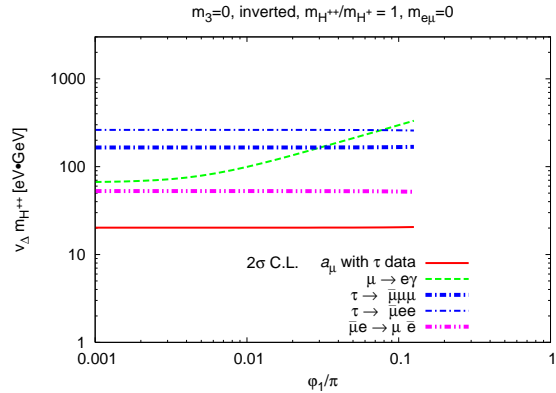}
\vspace*{-10mm}
\caption{
 The $\varphi_1$-dependence of lower bounds on $\vm$
for cases of $m_{e\mu}=0$
with $m_3 = 0$ in the inverted mass ordering.
 $\theta_{13}^\magic$ and $\delta^\magic$ for $m_{e\mu}=0$
are functions of $\varphi_1$ (See Appendix).
}
\label{fig:mphase_hem0_i}
\end{center}
\end{figure}
 Similarly to Fig.~\ref{fig:hem0_n},
lower bounds on $v_\Delta m_{H^{\pm\pm}}$
are shown in Fig.~\ref{fig:hem0_i}
for the inverted mass ordering.
 Note that $\varphi_1=\pi$ can not give $m_{e\mu}=0$
for the mass ordering because $s_{13}^\magic$ becomes too large.
 Very roughly speaking,
the results in Figs.~\ref{fig:hem0_i}(a) and (b)
are the same as those in Figs.~\ref{fig:hem0_n}(a) and (b),
respectively.
 A difference is that $\tmee$ can be
the most stringent bound at $m_3=0$
while it is not the case for $m_1=0$ in the normal mass ordering.
 Figure~\ref{fig:mphase_hem0_i} shows
the $\varphi_1$-dependence for $m_{e\mu}=0$ with $m_3 = 0$
where the $\varphi_2$-dependence vanishes.
 For $\varphi_1 \gtrsim 0.1\pi$,
$\sin^2{2\theta_{13}^\magic}$ becomes larger than 0.14.
 We see that $\meg$ can give the most stringent bound
even for the inverted mass ordering.
 For $m_3 = 0.2\,\eV$,
Fig.~\ref{fig:mphase_hem0_n}(b) is almost applicable
to see which process gives the most stringent bound.

\begin{figure}[t]
\begin{center}
\includegraphics[origin=c, angle=-90, scale=0.3]
{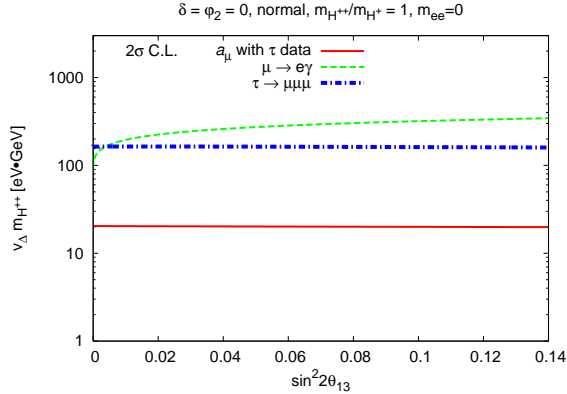}
\vspace*{-10mm}
\caption{ Lower bounds on $\vm$
for cases of $m_{ee}=0$ with $\delta = \varphi_2 = 0$
in the normal mass ordering.
 Values of $m_1^\magic$ and $\varphi_1^\magic$ are given
by the condition $m_{ee}=0$
depending on $\theta_{13}$ (See Appendix).
}
\label{fig:hee0}
\end{center}
\end{figure}
 Figure~\ref{fig:hee0} is the result for the case of $m_{ee}=0$
which is possible only in the normal mass ordering.
 Formulae of the magic values $m_1^\magic$ and $\varphi_1^\magic$
for $m_{ee}=0$ are shown in Appendix.
 The bound from $\mu \to e\gamma$ is the most stringent one
except for $\sin^2{2\theta_{13}} \lesssim 0.04$
where the bound from $\tau \to \bar{\mu}\mu\mu$ becomes
stronger than that.
 This is also the case for different values of
$\delta$ and $\varphi_2$.

\section{Conclusion}
 In the HTM,
it is impossible to have a contribution
to the muon anomalous MDM with a plus sign.
 Therefore,
the HTM is qualitatively disfavored
by positive $\Delta a_\mu[\tau]$ ($\Delta a_\mu[e^+e^-]$)
at $1.8\sigma$ ($3.7\sigma$).
 If we deal with $2\sigma$ bounds to
avoid the disagreement with $\Delta a_\mu[\tau]$,
$\vm$ must be greater than $10^{3}\,\eVGeV$
in most of parameter space of the HTM
in order to satisfy a strong constraint
on $\BR(\meee)$.
 We found that the bound on $\vm$ from $\tmee$ becomes
more stringent than that from $\meee$
in a region of $m_1 \gtrsim 0.06\,\eV$,
$\varphi_1 \lesssim 0.002\pi$,
$0.5\pi \lesssim \varphi_2 \lesssim 1.5\pi$,
and $\sin^2{2\theta_{13}} \lesssim 10^{-5}$.
 The bound from $\meee$ can be evaded
in cases of $m_{e\mu} = 0$~\cite{Chun:2003ej}
and $m_{ee} = 0$~\cite{Akeroyd:2009nu}.
 In the case of $m_{e\mu} = 0$
in the normal mass ordering,
the strongest bound is given by $\meg$
for $m_1 \lesssim 0.01\,\eV$ or $\varphi_1 \gtrsim 0.1\pi$
and by $\tau$ decays (mainly $\tmee$)
for $m_1 \gtrsim 0.01\,\eV$ with $\varphi_1 \lesssim 0.1\pi$.
 On the other hand,
$\tmee$ gives the strongest bound
in the normal mass ordering
except for $m_3 \simeq 0$ with $\varphi_1 \simeq 0.1\pi$
where $\meg$ gives the bound.
 For both of the mass orderings with $m_{e\mu}=0$,
the muonium conversion
gives the most stringent bound
if Majorana phases satisfy
$\varphi_2 \simeq 0.68\,\varphi_1$
for $\varphi_1, \varphi_2 \ll 1$ and $m_1 \gtrsim 0.1\,\eV$.
 In the case of $m_{ee} = 0$,
the strongest bound is obtained from $\meg$
except for the case of $\sin^2{2\theta_{13}}\lesssim 0.04$
where the bound is given by $\tmmm$.
 By looking over all cases,
we see that $\vm \gtrsim 150\,\eVGeV$ should be satisfied
in the HTM\@.
 If $m_{H{\pm\pm}}$ is measured,
the bound can be the lower bound on $v_\Delta$
though there remains a possibility of $v_\Delta = 0$
for which we can not use the correlation of $h_{\ell\ell^\prime}$
with $\sqrt{2} v_\Delta h_{\ell\ell^\prime} = m_{\ell\ell^\prime}$.

\section*{Acknowledgements}
 We thank A.G.~Akeroyd for useful discussions.
 The work of T.~F.\ is supported in part by the Grant-in-Aid
for Scientific Research from the Ministry of Education,
Science and Culture of Japan (No.~20540282).

\appendix
\section*{Appendix}

 We obtained formulae of the magic values
of $\theta_{13}$ and $\delta$,
which give $m_{e\mu} = 0$, as
\begin{eqnarray}
\cos\delta^\magic
&=&
 -
 \frac{
       c_{12} s_{12} c_{23} C
      }
      {
       s_{23} s_{13}^\magic
       \left\{
        m_3^2
        - s_{12}^4 m_2^2 \sin^2\varphi_1
        - ( s_{12}^2 m_2 \cos\varphi_1 + c_{12}^2 m_1 )^2
       \right\}
      },
\label{eq:cdmgc}\\
\sin\delta^\magic
&=&
 \frac{
       c_{12} s_{12} c_{23} D
      }
      {
       s_{23} s_{13}^\magic
       \left\{
        m_3^2
        - s_{12}^4 m_2^2 \sin^2\varphi_1
        - ( s_{12}^2 m_2 \cos\varphi_1 + c_{12}^2 m_1 )^2
       \right\}
      },
\label{eq:sdmgc}\\
s_{13}^\magic
&=&
 \frac{ c_{12} s_{12} c_{23} \sqrt{ C^2 + D^2 } }
      {
       s_{23}
       \left|
        m_3^2
        - s_{12}^4 m_2^2 \sin^2\varphi_1
        - ( s_{12}^2 m_2 \cos\varphi_1 + c_{12}^2 m_1 )^2
       \right|
      },
\label{eq:s13mgc}\\[5mm]
C
&\equiv&
 - c_{12}^2 m_1^2
 + s_{12}^2 m_2^2
 + m_1 m_2 \cos{2\theta_{12}} \cos\varphi_1
\nonumber\\
&&\hspace*{20mm}
{}- m_1 m_3 \cos\varphi_2
  + m_2 m_3 \cos(\varphi_1-\varphi_2),\\
D
&\equiv&
 - m_1 m_2 \sin\varphi_1
 + m_1 m_3 \sin\varphi_2
 + m_2 m_3 \sin(\varphi_1-\varphi_2),
\end{eqnarray}
where we define $s_{13}^\magic \equiv \sin{\theta_{13}^\magic}$.
 Note that $s_{13}^\magic \neq 0$ because
it requires $m_2 = m_1$.
 Note also that $s_{13}^\magic$ is not always acceptable;
 For example,
$\varphi_1=\pi$ in the inverted mass ordering
gives $s_{13}^\magic > 1$.
 These results are consistent with
$s_{13}^\magic$ and $\delta^\magic$
used in \cite{Chun:2003ej,Akeroyd:2009nu}
for $\varphi_1, \varphi_2 = 0\ \text{or}\ \pi$.
 Although mixing matrix in \cite{Akeroyd:2009nu} is defined as
$\nu_\ell = \sum_i U_{\ell i}^\ast \nu_i$
in stead of $\nu_\ell = \sum_i U_{\ell i} \nu_i$
used in this article,
there is no change in formulae of magic values
because the difference appears just as
the simultaneous flip of signs of all phases.

 On the other hand,
$m_{ee}=0$ can be achieved
only in the normal mass ordering.
 The magic values of $\varphi_1$ and $m_1$
for $m_{ee}=0$ are given
as functions of $s_{13}$ and $\varphi_2 - 2\delta$%
~\cite{Akeroyd:2009nu} by
\begin{eqnarray}
\sin\varphi_1^\magic
&\equiv&
 - \frac{ \sqrt{ (m_1^\magic)^2 + \Delta m^2_{31} } }
        { \ s_{12}^2 \sqrt{ (m_1^\magic)^2 + \Delta m^2_{21} } \ }\,
   t_{13}^2 \sin(\varphi_2 - 2\delta), \ \ \
\cos\varphi_1^\magic \leq 0,
\label{eq:phi1mgc}
\\
(m_1^\magic)^2
&\equiv&
 \frac{1}{ \cos^2{2\theta_{12}}
           - 2 \left(
                s_{12}^4 + c_{12}^4 \cos{2(\varphi_2 - 2\delta)}
               \right) t_{13}^4
           + t_{13}^8 }
\nonumber\\
&&\hspace*{0mm}
\times
  \Biggl[ \
   s_{12}^4 \cos{2\theta_{12}} \Delta m^2_{21}
   + \left\{
      s_{12}^4 \Delta m^2_{21}
      + \left(
         s_{12}^4 + c_{12}^4 \cos{2(\varphi_2 - 2\delta)}
        \right) \Delta m^2_{31}
     \right\} t_{13}^4
\nonumber\\
&&\hspace*{40mm}
 {}- \Delta m^2_{31} t_{13}^8
 {}- 2 c_{12}^2 t_{13}^2 \cos(\varphi_2 - 2\delta)
     \sqrt{ A + B\,t_{13}^4 } \
  \Biggr],
\label{eq:m1mgc}
\\
A
&\equiv&
 \left(
  s_{12}^4 \Delta m^2_{21}
  + \cos{2\theta_{12}} \Delta m^2_{31} 
 \right)
 s_{12}^4 \Delta m^2_{21},\\
B
&\equiv&
 \Bigl\{
  ( s_{12}^4 - c_{12}^4 \sin^2(\varphi_2 - 2\delta) )
  \Delta m^2_{31}
  - s_{12}^4 \Delta m^2_{21}
 \Bigr\} \Delta m^2_{31},
\end{eqnarray}
where we define $t_{13}\equiv s_{13}/c_{13}$.


\end{document}